\documentclass[a4paper,12pt]{article}
\usepackage{amsmath,amssymb,amsthm}
\usepackage{latexsym,graphicx,color,bbm,subfigure}
\usepackage{pdfsync}   
\usepackage{bbm}

\newcommand{\be}{\begin{equation}}
\newcommand{\ee}{\end{equation}}
\newcommand{\beq}{\begin{eqnarray}}
\newcommand{\eeq}{\end{eqnarray}}

\newcommand{\Tr}{{\rm Tr}}

\newcommand{\bea}{\begin{eqnarray}}
\newcommand{\eea}{\end{eqnarray}}
\def\Tr{ \hbox{\rm Tr}}

\makeatletter \@addtoreset{equation}{section} \makeatother

\begin{document}

\thispagestyle{empty}
\begin{flushright}
IFUP-TH/2013-06
\end{flushright}
\vspace{10mm}
\begin{center}
{\Large   \bf CONFINEMENT VIA STRONGLY-COUPLED  \\  NONABELIAN MONOPOLES} \footnote{Contribution to
 SCGT12 "KMI-GCOE Workshop on Strong Coupling Gauge
Theories in the LHC Perspective",   4-7 Dec. 2012, Nagoya University.}
\\[15mm]
{\large   Kenichi Konishi$^{a,b}$  
} \footnote{\it e-mail address: konishi(at)df.unipi.it.  
   }
\vskip 6mm
 
 $^a$  
~Department of Physics ``E. Fermi'', University of Pisa, \\
Largo Pontecorvo, 3, Ed. C, 56127 Pisa, Italy
\\
$^b$
  INFN, Sezione di Pisa,
Largo Pontecorvo, 3, Ed. C, 56127 Pisa, Italy

\vskip 6 mm

\bigskip
\bigskip

{\bf Abstract}\\[5mm]
{\parbox{14cm}{\hspace{5mm}
\small
  New types of confinement phase emerge as singular SCFT's appearing as infrared-fixed-points of ${\cal N}=2$ supersymmetric QCD (SQCD) are perturbed by an ${\cal N}=1$ adjoint mass term.  Based on a recent remarkable work on infrared-fixed-point SCFT of highest criticalities  by Gaiotto, Seiberg and Tachikawa,  we discuss physics of certain confining systems in $SU(N)$, $USp(2N)$ or  $SO(N)$ gauge theories.  These  show features different from  a straightforward dual superconductivity picture of confinement \`a la 't Hooft and Mandelstam,  which might suggest a new venue in exploring the quark confinement mechanism in the real-world  QCD. 
}
}
\end{center}
\newpage
\pagenumbering{arabic}
\setcounter{page}{1}
\setcounter{footnote}{0}
\renewcommand{\thefootnote}{\arabic{footnote}}


%


%






\section{Quark confinement versus chiral symmetry breaking}\label{aba:sec1}

Often a theory at an UV conformal fixed point flows in the infrared to 
another, infrared-fixed-point conformal theory.  If a small relevant operator  is either introduced by hand or generated dynamically,  the system can  instead  flow into one in confinement phase. If in some sense the relevant deformation is small, the knowledge about the infrared conformal theory (in the absence of such a  deformation) is important,  as the degrees of freedom in the latter  describe also how confinement and dynamical symmetry breaking take place. 

In Quantum Chromodynamics (QCD), the UV degrees of freedom are quarks and gluons, and their behavior at short distances is well understood because of asymptotic freedom, but  the nature of their collective behavior in the infrared is still covered by  mysteries.  The familiar  Nambu-'t Hooft-Mandelstam picture \cite{TH,NM} of confinement  assumes a dynamical gauge symmetry breaking, 
\be  SU(3) \to   U(1) \times U(1)   \to  {\bf 1} \;,  
\ee
with the magnetic monopoles $M_{i}\,\,$  ($i=1,2$) generated by the first step of gauge symmetry breaking,  acting as confinement order parameters.  Indeed,  their condensation $\langle M_{i} \rangle\ne 0$  would lead to a (dual) superconductor vacuum, so that the color electric flux is squeezed into narrow dual Abrikosov-Nielsen-Olesen (ANO) vortices: the quarks are confined, just as the magnetic monopoles of the usual electromagnetism would be in the standard type II superconductor. 

This well-known picture of confinement leads however to difficulties. If the confinement and chiral symmetry breaking take place simultaneously, as suggested by $SU(3)$ lattice simulation data, 
it is natural to assume that one or both of the magnetic monopoles carry flavor $SU_{L}(N_{f}) \times SU_{R}(N_{f})$ charges \`a la Jackiw-Rebbi \cite{JR};  their condensation leads both to confinement and chiral symmetry breaking. In softly broken ${\cal N}=2$  SQCD with $SU(2)$ gauge group and with $N_{f}=1,2,3$ flavors, exactly such a phenomenon is  dynamically realized at low energies \cite{SW1,SW2,APS,Others,CKM}.  
In the case of non-supersymmetric  QCD, however,  the pattern of the chiral symmetry breaking
\be   SU_{L}(N_{f}) \times SU_{R}(N_{f}) \to SU_{V}(N_{f})    \label{vector}  
\ee
would require one (or both) of the magnetic monopoles to carry both left and right  flavor $SU(N_{f}) $ charges
\be        M^{a}_{b}, \qquad  a,b =1,2, \ldots N_{f}\;,  
\ee
so that the condensation 
\be    \langle M^{a}_{b} \rangle = c\, \delta^{a}_{b}\, \Lambda 
\ee
explains the symmetry breaking pattern Eq.~(\ref{vector}).   This scenario however implies an accidental global $SU(N_{f}^{2})$ 
symmetry of the Abelian monopole theory  and as a consequence, a proliferation of Nambu-Goldstone bosons  (in the realistic case of $N_{f}=2$,  twelve  NG bosons instead of three), of which of course there are no traces in nature. 

Another, perhaps more serious,  problem is the fact that the low-energy theory has 
\be    \Pi_{1}(U(1) \times U(1)) = {\mathbbm Z}\times  {\mathbbm Z},
\ee
implying a doubling of the meson spectrum (quarks confined by two distinct confining strings), while in the real world one observes unique, universal $q- {\bar q}$ meson trajectory for each flavor quantum number. 

A possible way out of this conundrum  is that the system does not completely Abelianize, i.e., that  the gauge symmetry dynamically breaks as 
\be  SU(3) \to   SU(2) \times U(1)   \to  {\bf 1} \;. \label{NAbreaking}
\ee
In this case magnetic monopoles generated are of nonAbelian variety.   As 
\be    \Pi_{1}(SU(2) \times U(1)) = {\mathbbm Z},
\ee
there would be a unique confining string in this case.  We shall not discuss here the well-known ``existence problems''  for the nonAbelian monopoles, but it is believed that the light flavors play a crucial role for the quantum mechanical behavior of the nonAbelian monopoles \cite{KK} solving the difficulties arising for the monopoles in pure Yang-Mills theories.    It is to be seen whether and how the problem of the proliferation of the Nambu-Goldstone bosons can be avoided in this case.


At the same time, however,  this introduces a new (apparent)  difficulty.  In contrast to what happens in the $r$ vacua of the  softly broken ${\cal N}=2$ supersymmetric QCD \cite{CKM}   (see below), the nonAbelian monopoles of the standard QCD are expected to be strongly coupled at low energies. One might wonder whether it makes sense to assume that the strongly coupled quarks and gluons get replaced in the infrared by another strongly coupled (e.g., magnetic) system. Actually,  the idea is not new:  there are many known examples of systems possessing similar properties (tumbling gauge theories \cite{Raby}, duality cascades \cite{Klebanov}, etc.) and, as will be seen below, some supersymmetric version of QCD  exhibit exactly this kind of behavior. 

It is possible that ultimately one must accept the idea that the color magnetic degrees of freedom of QCD are strongly coupled, and that confinement and dynamical chiral symmetry breaking are phenomena subtler than expected from a straightforward dual superconductivity idea.

\section{What softly broken  ${\cal N}=2$ SQCD  teaches  us}  

As many exact quantum results are known in the softly broken  ${\cal N}=2$   SQCD,  it is natural to ask what these theories can tell us about confinement, dynamical symmetry breaking and about the mechanism underlying them.  Confining vacua in these theories can be classified
as follows.

\begin{table}[b]
\begin{center}
\vskip .3cm
\begin{tabular}{|ccccccc|}
\hline
&   $SU(r)  $     &     $U(1)_0$    &      $ U(1)_1$
&     $\ldots $      &   $U(1)_{N-r-1}$    &  $ U(1)_B  $  \\
\hline
$n_f \times  {\cal M}$     &    ${\underline {\bf r}} $    &     $1$
&     $0$
&      $\ldots$      &     $0$             &    $0$      \\ \hline
$M_1$                 & ${\underline {\bf 1} } $       &    0
&
1      & \ldots             &  $0$                   &  $0$  \\ \hline
$\vdots $  &    $\vdots   $         &   $\vdots   $        &    $\vdots   $
&             $\ddots $     &     $\vdots   $        &     $\vdots   $
\\ \hline
$M_{N-r-1} $    &  ${\underline {\bf 1}} $    & 0                     & 0
&      $ \ldots  $            & 1                 &  0 \\ \hline
\end{tabular}
\caption{\footnotesize The massless non-Abelian and Abelian monopoles  and their charges  at the $r$ vacua
at the root of a ``non-baryonic'' $r$-th   Higgs branch.  }
\label{aba:tbl1}
\end{center}
\end{table}

\begin{figure}
\begin{center}
\includegraphics[width=6in]{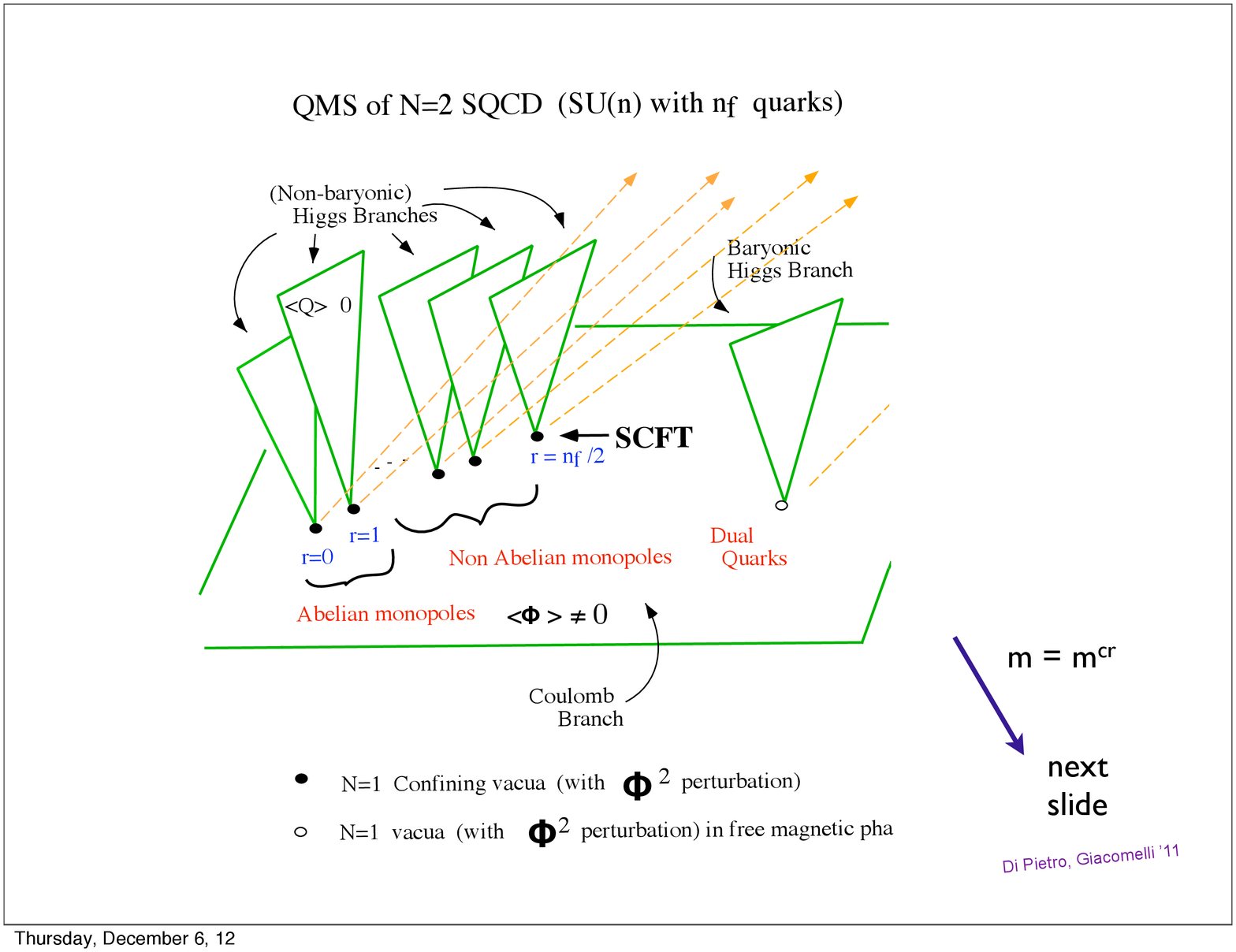}
\caption{\footnotesize  A schematic view of the quantum moduli space of vacua in softly broken ${\cal N}=2$  SQCD with $SU(N)$ gauge group and $N_{f}$ flavors of quarks. Upon $\mu \Phi^{2}$ perturbation only the interceptions where  various Higgs branches ($\langle Q \rangle \ne 0$) and the Coulomb branch  ($\Phi \ne 0$) meet,  survive (the black and while points).  The black points are the $r$ vacua; the white point  represents the baryonic Higgs branch root with the vacua which are not confining. In the limit $m \to m^{cr}$ the r-vacua collapse to a singular SCFT as in the $USp(2N)$ case  (the next Figure).  }
\label{sunQMS}
\end{center}
\end{figure}
\begin{figure}
\begin{center}
\includegraphics[width=6in]{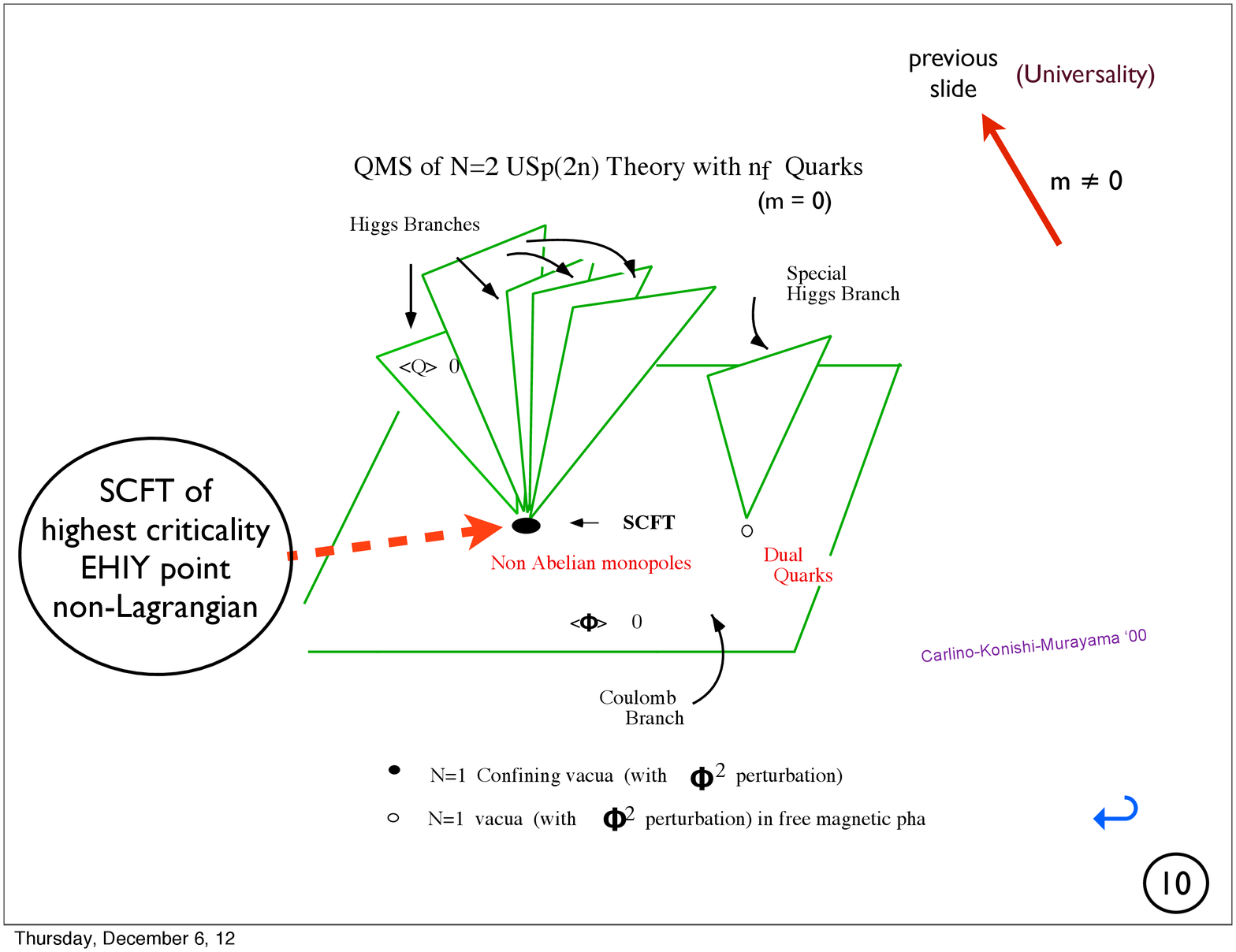}
\caption{\footnotesize The quantum moduli space of vacua in $USp(2N)$ theory with $N_{f}$ flavors of quarks. 
For vanishing bare quark masses  all confining vacua are in the Chebyshev points (the black point) which are SCFT of the highest criticality.  Upon turning on the equal bare quark masses  $m \ne 0$, they split into various $r$ vacua 
which are identical to those in the $SU(N)$ theory (the previous Figure). }
\label{uspQMS}
\end{center}
\end{figure}

\begin{description}

  \item[(i)]  Abelian dual superconductor vacuum   is realized in certain class of theories, as in the $SU(2)$ gauge theories, or in all pure ${\cal N}=2$ supersymmetric theories with any gauge group $G$.   The system dynamically Abelianizes, and in low energies becomes a $U(1)^{R}$ theory, where $R$ is the rank of the gauge group $G$. In the case of $SU(2)$ theory with $N_{f}=1,2,3$ flavors, the monopole condensation induces confinement and 
dynamical flavor symmetry breaking.  This is beautiful, but does not look like what happens in the real-world QCD, as discussed above.  
  
  \item[(ii)]  Massless nonAbelian monopoles, carrying flavor quantum numbers appear in the so-called $r$-vacua of  SQCD.  (See Fig. \ref{sunQMS}).  In the $SU(N)$ theory
  with $N_{f}$ flavors, the low-energy action is \cite{APS,CKM} an $SU(r)\times U(1)\times \ldots U(1)$  theory,  with light monopoles carrying the quantum numbers   shown in Table~\ref{aba:tbl1}.   The low-enrgy effective $SU(r) \times U(1)$ theory is a local, infrared-free theory: the monopoles are weakly coupled  \footnote{The nonAbelian monopoles acquire flavor  multiplicities through the fermion zeromodes \cite{JR}:   this is essential  for the sign flip of the beta function.
   Indeed the quantum $r$ vacua occur only for $r \le N_{f}/2$.}. 
  Upon $\mu \Phi^{2}$ perturbation the monopoles condense in a dual-color-flavor locked vacuum, 
  \be      \langle M_{i}^{\alpha} \rangle =  \delta_{i}^{\alpha} \, \sqrt{\mu \Lambda}\;,    
  \ee
  breaking the symmetry as 
\be    SU(N_{f}) \times U(1) \to   U(r) \times U(N_{f}-r)\;.  
\ee
Again, in spite of the beautiful aspects  -  nonAbelian monopoles appear  quantum mechanically  as low-energy degrees of freedom, and they act as order parameters of confinement and dynamical symmetry breaking  -
these systems do not look like a good model for QCD.

  \item[(iii)]  Still another possibility, realized in softly broken ${\cal N}=2$  supersymmetric  theories,    is that strongly interacting nonAbelian monopoles appear in a nontrivial infrared-fixed point SCFT. The low-energy effective theory is a non-Langrangian theory involving monopoles and dyons.  This occurs  in  $USp(2N)$ and $SO(N)$ theories with vanishing bare quark masses \cite{CKM}
   (see Fig. \ref{uspQMS}),  and as recently realized  by Di Pietro and Giacomelli \cite{DiPGiac},  for some critical value of the quark masses,  in $SU(N)$ theories as well.   
 Our discussions below will be mainly concerned with these singular vacua.

\end{description}

\section{Recent developments} 

A development of considerable interest for our purpose  is the recent elucidation of the nature of certain infrared-fixed-point SCFT of highest criticalities \cite{GST, Simone}  in  the context of ${\cal N}=2$ supersymmetric QCD.   These SCFT's occur at particular points of the vacuum moduli space and/or for special values  of the bare quark mass parameters.  A straightforward interpretation of these points  would involve
 monopoles and dyons in an infinite-coupling regime, where it is not easy draw any clear physics picture.   
 

    
\subsection{Argyres-Seiberg $S$ duality}

The first key step forward has been the discovery  by  Argyres and Seiberg \cite{AS}   of an elegant  S-dual description of some  ``infinitely-strongly-coupled  SCFT''.   
   For example consider the  ${\cal N}=2$, $SU(3)$ gauge theory with $N_{f}=6$ flavors of hypermultiplets (quarks).  This theory is superconformal, with an exactly marginal coupling constant  $g$. By studying the behavior of the Seiberg-Witten curve 
in the limit $g \to \infty$,  Argyres and Seiberg were able to show that the system in this limit admits a weakly coupled dual description. 
  The dual is two separate SCFT sectors coupled weakly by  $SU(2)$ gauge interactions.   
In this particular $SU(3)$ theory, one sector is a free doublet of hypermultiplet, whereas the other sector is a non-Lagrangian SCFT with $E_{6}$ global symmetry \cite{MN}, whose $SU(2)$ subgroup ($SU(2) \times SU(6) \subset E_{6}$) is weakly gauged. The commutant $SU(6)$ and the $U(1)$ associated wit the free doublet  hypermultiplet make up the $U(6)$ global symmetry of the underlying $SU(3)$, $N_{f}=6$, theory.   

In another example of  $USp(4)$ theory with $N_{f}=6$  (with a global $SO(12)$ symmetry),  the infinite strong coupling limit is dual to an SCFT  with $E_{7}$ global symmetry,  whose  $SU(2)$ subgroup  ($SU(2) \times SO(10) \subset E_{7}$) is weakly gauged.

\subsection{Gaiotto-Seiberg-Tachikawa duals}

The next, crucial step was made by  Gaiotto, Seiberg and Tachikawa  (GST) \cite{GST}, who applied the  Argyres-Seiberg S-dual description to those SCFT's   appearing as infrared fixed points of  ${\cal N}=2$  $SU(N)$  SQCD.  This way they have been able to solve certain puzzles which plagued  earlier studies on the IFPT conformal theories of highest criticalities. 

In the case of $SU(N)$ gauge theory with $N_{f}=2n$ flavors,  the Seiberg-Witten (SW) curve is given by 
\be    y^{2}= (x^{N}+ u_{1} x^{N-1}+ u_{2} x^{N-2}+\ldots +  u_{N})^{2}   -  \Lambda^{2N- 2n}\, \prod_{i=1}^{2n}  (x + m_{i})
\label{SWc} \ee
Setting $u_{j}=m_{i}=0$ (except for $u_{N-n} $ which is chosen at $u_{N-n} = \Lambda^{N-n}$)   one is at the EHIY point \cite{EHIY}  where 
\be   y^{2} \sim  x^{N+n} \;. 
\ee
A straightforward scaling argument on the fluctuation around this point, by requiring the meromorphic (SW) differential
\be   \lambda \sim  \frac{ y\, dx}{ x^{n}}    
\ee
to have the unit canonical dimension,  would assign the dimensions  $[x] =\tfrac{1}{N+1}$; $\,[y]=\tfrac{N+n}{2(N+1)}$.  Assumption that the curve scales uniformly would then determine how various $u_{i}$'s and masses scale.  But this leads to mass parameters associated with nonAbelian flavor symmetries to possess anomalous scaling dimensions, which they should not  \cite{APSW}.  
This is the same,  well known argument that nonAbelian flavor charges do not get renormalized, as they satisfy inhomogeous current algebra commutation relations.  
From the point of view of finding infrared SCFT as points of quantum moduli space of vacua where the SW curves exhibit certain singular behaviors, 
 there is an {\it a priori}  ambiguity how to define (i.e., how to approach) the conformal limit,  $u_{i}\to 0,$  $ m_{i} \to 0$.    

\begin{figure}
\begin{center}
\includegraphics[width=5in]{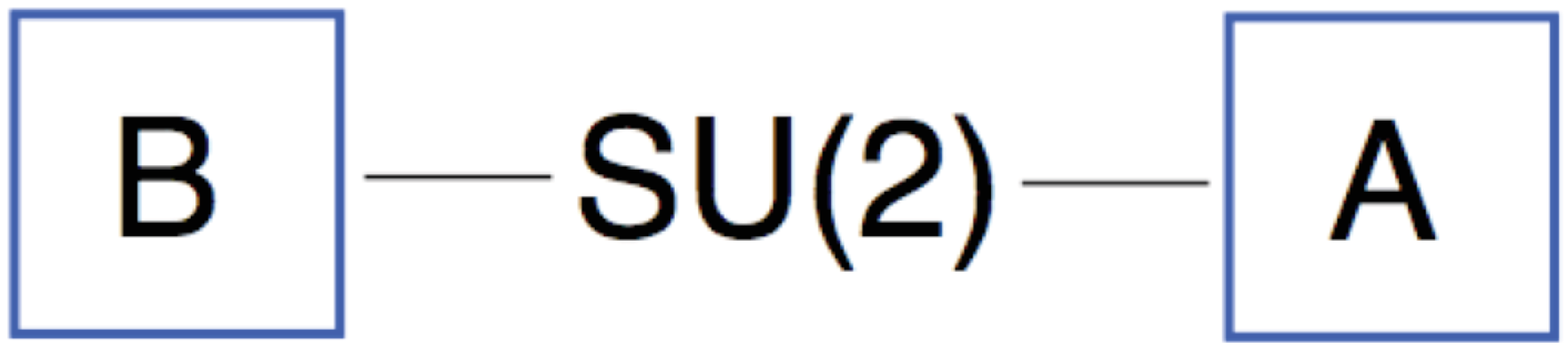}
\caption{ }
\label{GST}
\end{center}
\end{figure}

It was found \cite{GST} that in order to have an  SCFT with the correct mass dimensions it is necessary to scale various  $u_{i}$'s  towards  $0$   nontrivially,  as   
\be    u_{N-n+2}\sim O(\epsilon_{A}^{2}), \quad  u_{N-n+3}\sim O(\epsilon_{A}^{3}), \quad\ldots  \quad  u_{N}\sim O(\epsilon_{A}^{n}), \quad  
\ee 
\be    u_{1}\sim O(\epsilon_{B}), \quad  u_{2}\sim O(\epsilon_{B}^{2}), \quad \ldots  \quad  u_{N-n+2}\sim O(\epsilon_{B}^{N-n+2}), \quad  
\ee 
where  
\be  \epsilon_{A}^{2} \sim \epsilon_{B}^{N-2+2}, \qquad   \epsilon_{A} \ll  \epsilon_{B}
\ee
and 
\be   m_{i_{1}}\cdots m_{i_{k}} \sim  \epsilon_{A}^{k}\;.   
\ee
Accordingly,  the branch points of the SW curve (\ref{SWc}) get separated into two groups of different orders of magnitudes
\be    x \sim \epsilon_{A}, \qquad {\rm and} \qquad  x  \sim \epsilon_{B}
\ee 
and consequenty the system splits into different  sectors. The result is the system composed of 
 \begin{description}
  \item[(i)]   Some decoupled $U(1)^{N-n-1}$  gauge multiplets;
  \item[(ii)] An SU(2)   gauge multiplet (infrared free) coupled to the $SU(2)$ flavor symmetries of the two SCFT's, A and B;
  
  \item[(iii)]  The A sector is a SCFT entering the Argyres-Seiberg dual of  $SU(n)$, $N_{f}= 2n$ theory, having $SU(2) \times SU(2n)$  flavor symmetry;  
  \item[(iv)]  The B sector is a maximally singular SCFT of the $SU(N-n+1)$ theory with two flavors, 
\end{description}
 which may be schematically represented as in Fig.~\ref{GST}.
 
{\noindent \bf Remarks: }
 \begin{description}
  \item[(i)]   In the case $n=2$, the $A$ theory describes simply three free hypermultiplets;    for $n=3$,  it is a non-Lagrangian SCFT with $E_{6}$ global symmetry first found by Minahan and Nemeschansky \cite{MN};  
  \item[(ii)]   For $N=3$, $n=2$,  the $B$ sector corresponds to the most singular SCFT of the $SU(2)$, $N_{f}=2$ theory \cite{APSW}.
   \item[(iii)]   The Gaiotto-Seiberg-Tachikawa analysis has been generalized to the cases of  $USp(2N)$ and $SO(N)$ theories by  Giacomelli \cite{Simone}.  
 \end{description}

   These developments enable us to study new types of confining systems arising as deformation of these strongly critical SCFT's, to which we now turn
   our attention.

\section{Confinement near the singular vacua} 

\subsection{$USp(2N)$  theory with  $N_{f}=2n$}

In the ${\cal N}=2$  $USp(2N)$ theory, the relevant most singular  vacua are \cite{CKM}  two ``Chebyshev''  vacua,   at    $\phi_{1}=\phi_{2}=\ldots =\phi_{n-1}=0$,    $\phi_{n}^{2}= \pm \Lambda^{2}$  (other $\phi$'s are  determined \`a la Douglas-Shenker by using a Chebyshev polynomial):   
\be  x\, y^{2}  \sim    [\, x^{n} (x- \phi_{n}^{2})\,]^{2} - 4\, \Lambda^{4}\, x^{2n} =   x^{2n} \, ( x-\phi_{n}^{2} - 2 \, \Lambda^{2}) \, ( x-\phi_{n}^{2} +  2 \, \Lambda^{2}) 
\label{Cheby}\ee
that is, 
\be    y^{2} \sim x^{2n}\;   \label{Chebysing}
\ee
(see Fig.~\ref{uspQMS}). 
Such a singular behavior implies that  strongly-coupled massless monopoles and dyons, relatively nonlocal to each other, appear simultaneously in the low-energy effective action.

The strategy adopted in Carlino et. al. \cite{CKM} was to try to  ``resolve'' this vacuum, by introducing generic, nearly equal quark masses $m_{i}$ alongside the adjoint scalar mass $\mu$.   
By requiring the factorization property of the SW curve to be of maximally Abelian type (the criterion for ${\cal N}=1$ supersymmetric vacua),   this point was found to split into  various $r$ vacua  which are local $SU(r)\times U(1)^{N-r}$ gauge theories, identical to those appearing in the infrared limit of   $SU(N)$  SQCD (the universality of the infrared fixed points).  One type of the Chebyshev vacua ($\phi_{n}^{2}=+  2\Lambda^{2}$)  yields 
\be    \binom{N_{f}}{0} + \binom{N_{f}}{2} + \ldots \binom{N_{f}}{N_{f}} =  2^{N_{f}-1}    \label{one}
\ee
vacua,  whereas the other vacua  ($\phi_{n}^{2}=- 2\Lambda^{2}$)  split into odd $r$ vacua, with the total multiplicity
\be    \binom{N_{f}}{1} + \binom{N_{f}}{3} + \ldots \binom{N_{f}}{N_{f}-1} =  2^{N_{f}-1}\;.  \label{two}
\ee
Now the GST dual description of the point  (\ref{Cheby})     was found by Giacomelli  \cite{Simone}: 
 \begin{description}
  \item[(i)]   Some decoupled $U(1)^{N-n}$  gauge multiplets;
  \item[(ii)] An SU(2)   gauge multiplet (infrared free) coupled to the $SU(2)$ flavor symmetries of the two SCFT's, A and B;
  
  \item[(iii)]  The A sector is a non-Lagrangian SCFT having $SU(2) \times SO(4n)$  flavor symmetry;  
  \item[(iv)]  The B sector is a free doublet coupled to a $U(1)$ gauge field.
\end{description}
For a particular choice of the number of flavors, $N_{f}=4$, even the A sector becomes a trivial SCFT:    four free doublets, so let us 
concentrate on this particular case. 

We wish to verify that the SCFT vacua, {\it deformed by the $\mu \Phi^{2}$ perturbation},  are correctly described by the GST variables.   The superpotential is given by 
\beq  \label{vacmass} \sqrt{2} \, Q_{0} A_{D} {\tilde Q}^{0} +  \sqrt{2} \, Q_{0} \phi {\tilde Q}^{0} + \sum_{i=1}^{4}  \sqrt{2} \, Q_{i} \phi {\tilde Q}^{i}  +  \mu A_{D} \Lambda  +  {\mu}\,  \Tr \phi^{2}
+  \sum_{i=1}^{4}   m_{i}\, Q_{i} {\tilde Q}^{i}\;.       
\eeq
For equal and nonvanishing masses the system has $SU(4)\times U(1)$ flavor symmetry. In the massless limit the symmetry gets enhanced to $SO(8)$,
in accordance with the symmetry of the underlying $USp(2N)$  theory.    


The vacuum equations are:   
\beq    \sqrt{2} \, Q_{0} {\tilde Q}_{0} + \mu \Lambda =0\;;   \label{eq1}
\eeq
\beq   (\sqrt{2} \, \phi + A_{D})  {\tilde Q}_{0}= Q_{0} \, (\sqrt{2} \, \phi + A_{D}) =0\;;  \label{eq2}
\eeq
\beq  \sqrt{2} \,\,\left[\, \frac{1}{2}  \sum_{i=1}^{4}  Q_{i}^{a}   {\tilde Q}_{b}^{i} -  \frac{1}{4}  \delta_{b}^{a}  Q_{i}{\tilde Q}^{i} +  \frac{1}{2} Q_{0}^{a}{\tilde Q}^{0}_{b}- \frac{1}{4} \delta^{a}_{b}  Q_{0}{\tilde Q}^{0} \, \right] + \mu \, \phi^{a}_{b}=0\;;  \label{eq3}
\eeq
\beq    (\sqrt{2} \,\phi +m_{i} )\, {\tilde Q}^{i}=  Q_{i}\, (\sqrt{2} \,\phi + m_{i})  =0, \qquad \forall i\;.   \label{eq4}
\eeq
The first tells that $Q_{0}\ne 0$.  By gauge choice
\beq  Q_{0} =  {\tilde Q}_{0}= \left(\begin{array}{c}   2^{-1/4}\sqrt{-\mu \Lambda}  \\ 0 \end{array}\right)
\eeq
so that 
\beq    \frac{1}{2} Q_{0}^{a}{\tilde Q}^{0}_{b}- \frac{1}{4}  (Q_{0}{\tilde Q}^{0}) \, \delta^{a}_{b} = \frac{(-\mu \Lambda)}{4 \sqrt{2}} \, \tau^{3}\;.  
\eeq
The second equation can be satisfied by adjusting $A_{D}$.


The  solutions can be found by having one of $Q_{i}$'s canceling the contributions of   $Q_{0}$ and $\phi$ in   Eq~(\ref{eq3}).  Which of $Q_{i}$ is nonvanishing is related to the
value of $\phi$  through Eq~(\ref{eq4}).   For instance, four  solutions can be found by choosing  ($i=1,2,3,4$)
\beq   a=  -\frac{m_{i}}{\sqrt{2}}, \qquad   Q_{i}= {\tilde Q}_{i} = \left(\begin{array}{c}f_i \\0\end{array}\right) ;\qquad  Q_{j}= {\tilde Q}_{j}=0, \quad  j \ne i, 
  \label{sol11}  \eeq
such that  
\beq   f_{i}^{2}=    \frac{\mu \Lambda - 4\, a }{\sqrt{2}} =   \mu ( \frac{\Lambda}{\sqrt{2}} + 2 m_{i})\;.  
\eeq
There are four more solutions  with the similar form as above but with $a= +\tfrac{m_{i}}{\sqrt{2}}$ \cite{noi2New}.  
These are unrelated to each other and to (\ref{sol11}) by any  gauge transformation,  so that 
we find  $2^{3}=8$ solutions in all,  consistently with  Eq.~(\ref{one}).

Approaching the equal mass limit these  eight solutions group into two  set of four nearby vacua, clearly related by the $SU(4)$. 
So they are the $4+4 =8$,  two  $r=1$ vacua,  from one of the Chebyshev vacua, see Eq.~(\ref{two}).  The other Chebyshev vacuum should give $1 + 6 + 1=8$ vacua, corresponding to  $r=0,2$ vacua. Where are they?

A possible solution is that in the other Chebyshev vacuum  the superpotential has a similar form as (\ref{vacmass}) but with $Q_{i}$'s carrying different flavor charges.   The $SU(4)$ symmetry of the equal mass theory may be represented as  $SO(6)$:
\beq   \label{vacmassBis}   \sqrt{2} \,Q_{0} A_{D} {\tilde Q}^{0} +  \sqrt{2} \, Q_{0} \phi {\tilde Q}^{0} + \sum_{i=1}^{4}  \sqrt{2} \,Q_{i} \phi {\tilde Q}^{i}  +  \mu A_{D} \Lambda  +  {\mu}\,  \Tr \phi^{2}
+     \sum_{i=1}^{4}   {\tilde m}_{i}\, Q_{i} {\tilde Q}^{i}\;,  
\eeq
where      
\bea   {\tilde m}_{1} =    \frac{1}{4}  (m_{1}+m_{2}- m_{3}-m_{4})\;;  \nonumber \\   
{\tilde m}_{2}=    \frac{1}{4}  (m_{1}-m_{2}+ m_{3}-m_{4})\;;  \nonumber \\ 
  {\tilde m}_{3}=    \frac{1}{4}  (m_{1}-m_{2}- m_{3}+m_{4})\;;\nonumber \\ 
    {\tilde m}_{4}=  \frac{1}{4}  (m_{1}+m_{2} + m_{3} + m_{4})\;.    \label{masses}
\eea
The correct realization of the underlying symmetry in various cases is not obvious, so let us check them:   
\begin{description}
\item[(i)]  In the equal mass limit, $m_{i}=m_{0}$,  
\beq       {\tilde m}_{4}=m_{0}, \qquad    {\tilde m}_{2}= {\tilde m}_{3}= {\tilde m}_{4}=0\;, 
\eeq
so  the symmetry is
\beq   U(1)\times SO(6) = U(1)\times SU(4)\;.
\eeq
  Clearly in the $m_{i}=0$ limit the symmetry is enhanced to $SO(8)$. 
  \item[(ii)]   $m_{1}=m_{2}$,    $m_{3}$, $m_{4}$ generic.  
  In this case    ${\tilde m}_{2}= -  {\tilde m}_{3}$ and  ${\tilde m}_{4}$ and  ${\tilde m}_{1}$ are generic, so  the symmetry is   $U(1)\times U(1) \times U(2)$, as in the underlying theory;
  \item[(iii)] $m_{1}=m_{2} \ne 0$,    $m_{3}=m_{4}=0$.    In this case,  ${\tilde m}_{4}=  {\tilde m}_{1}\ne 0$ and ${\tilde m}_{2}=  {\tilde m}_{3} = 0$, so
  obviously the symmetry is $U(2)\times SO(4)$ both in the UV and in  (\ref{vacmassBis}).  
  
  \item[(iv)] $m_{1}=m_{2}=m_{3}\ne 0$,   $m_{4}$ generic.  In this case,  ${\tilde m}_{1}=  {\tilde m}_{2} =  -  {\tilde m}_{3}  \ne 0$,     ${\tilde m}_{4}$ generic. 
  Again the symmetry is $U(3)\times U(1)$ both at the UV and IR.

  \item[(v)]  $m_{1}=m_{2} \ne 0$  and   $m_{3}=m_{4} \ne 0$ but $m_{1}\ne m_{3}$.  In this case  $ {\tilde m}_{2} =  {\tilde m}_{3} = 0$ and  
  $ {\tilde m}_{4}$  and  ${\tilde m}_{1}$ generic.  The flavor symmetry is 
  \beq    SO(4) \times  U(1) \times U(1) =  SU(2)\times SU(2) \times  U(1) \times U(1)\;; 
  \eeq
  this is equal to the symmetry 
  \beq    (SU(2) \times U(1)) \times   (SU(2)\times U(1))
  \eeq
  of the underlying theory.
  \item[(vi)]    $m_{1}\ne 0$,  $m_{2}=m_{3}=m_{4}=0.$  In this case  ${\tilde m}_{1}={\tilde m}_{2}={\tilde m}_{3}={\tilde m}_{4}\ne 0$.
  The symmetry is $U(1)\times SO(6)$ in the UV, and  $U(4)$ in the   infrared. 
   \item[(vii)]    $m_{1}\ne 0$,  $m_{2}\ne 0,$  $ m_{1} \ne m_{2}$,  $m_{3}=m_{4}=0.$  In this case  ${\tilde m}_{1}={\tilde m}_{4}$   ${\tilde m}_{2}={\tilde m}_{3}\ne   {\tilde m}_{1}$.
  The symmetry is $U(1)^{2}\times SO(4)$ in the UV, and  $U(2)\times U(2)$ in the   infrared. 

\end{description}
The cases of masses equal except sign, e.g.,   $m_{1}= - m_{2}$,  are similar. 

Thus in all cases Eq.(\ref{vacmassBis}) has the correct symmetry properties as the underlying theory.  The vacuum solutions which follow from  it are similar to those
found from   Eq.(\ref{vacmass}),  with simple replacement, 
\beq   m_{i} \to {\tilde m}_{i}
\eeq
so  there are  $8$ of them.   The interpretation and their positions in the quantum moduli space (QMS) are different, however.    In the equal mass limit, $m_{i}\to m_{0}$,   The two solutions with 
\beq  a = - \frac{{\tilde m}_{4}}{\sqrt{2}}, \qquad {\rm or} \qquad    a =  \frac{{\tilde m}_{4}}{\sqrt{2}}, 
\eeq
can be regarded as  two  $r=0$ vacua. Note that as $|f_{1}| \ne |g_{1}|$ they are in distinct points of the moduli space.  
On the other hand, in the other six vacua  $a=0$ always and     $|f_{i}| = |g_{i}| $, these six solutions are at the same point of  
the moduli space: they can be associated with the $r=2$ (sextet) vacua.

\noindent{\bf Remarks:} 
%
The flavor charges (\ref{masses}) suggest that $Q$'s  are really non-Abelian magnetic monopoles, as semiclassically magnetic 
monopoles appear in the spinor representations of $SO(2N_{f})$.

\subsection{Confinement and dynamical symmetry breaking}

In the massless limit,  in the original SW description we had singular Chebyshev vacua.   Having relatively non-local,  strongly coupled monopoles and dyons, physics there was not very obvious, especially as to the effect of the $\mu \Phi^{2}$ perturbation and the types of confinement and dynamical symmetry breaking which might ensue. 

In the GST description, once can take smoothly the $m_{i} \to 0$ limit,  keeping $\mu \ne 0$,  to  get a system,  
\beq     \sqrt{2} \, Q_{0} A_{D} {\tilde Q}^{0} +  \sqrt{2} \, Q_{0} \phi {\tilde Q}^{0} + \sum_{i=1}^{4}  \sqrt{2} \, Q_{i} \phi {\tilde Q}^{i}  +  \mu A_{D} \Lambda  +  {\mu}\, \Tr \phi^{2}\;.
\eeq
The vacuum of this system can be easily found \cite{noi2}:
\beq  Q_{0} =  {\tilde Q}_{0}= \left(\begin{array}{c}   2^{-1/4}\sqrt{-\mu \Lambda}  \\ 0 \end{array}\right)
\eeq
\beq   \phi =0, \quad  A_{D}=0\;.  \label{sol2}
\eeq 
The contribution from $Q_{i}$'s must then cancel that of   $Q_{0}$ in  Eq.~(\ref{eq3}).   
By flavor rotation  the nonzero VEV can be attributed to $Q_{1}, {\tilde Q}^{1}$, i.e., either of the form
\beq     (Q_{1})^{1} =  ({\tilde Q}^{1})_{1} =2^{-1/4} \sqrt{\mu \Lambda} \;,    \qquad  Q_{i}={\tilde Q}_{i}=0, \quad i=2,3,4.    \label{sol3a}
\eeq
or  
\beq     (Q_{1})^{2} =  ({\tilde Q}^{1})_{2}   = 2^{-1/4}  \sqrt{-  \mu \Lambda} \;,    \qquad  Q_{i}={\tilde Q}_{i}=0, \quad i=2,3,4. \label{sol3b}
\eeq
 The $U(1)$ gauge symmetry is broken by the $Q_{0}$ condensation:  an ANO vortex is formed. As the gauge group of the underlying theory is  simply connected,
such a low-energy vortex must end. The quarks are  confined. The flavor symmetry breaking 
\beq 
  SO(8) \to   U(1) \times  SO(6) =   U(1) \times SU(4) = U(4),  \label{right}
\eeq
is induced by the condensation of  $Q_{1}$, which does not carry the $U(1)$ gauge charge.  We note that the pattern of the symmetry breaking above agrees with that found at large $\mu$  \cite{CKM}.

The vortex is made of the $Q_{0}$ field and the effective Abelian gauge
field.      The most interesting feature of this system is that there is no dynamical Abelianization, i.e., the effective low-energy gauge group is $SU(2)\times U(1)$.  
The confining string is unique and does not leads to the doubling of the meson spectrum \footnote{This last remark is true for $N=n$.   For $USp(2N)$ theories with $N > n=N_{f}/2$  there are decoupled $U(1)^{N-n}$ sectors. }. The global symmetry breaking of the low-energy effective theory is the 
right one (\ref{right}), but the vacuum is not color-flavor locked. The confining string is of Abelian type, and is not a non-Abelian vortex as the one appearing in an $r$ vacuum \cite{noi2}.  
These facts clearly  distinguish the confining system found here both from the standard Abelian dual superconductor type systems and from the 
non-Abelian dual Higgs system found in the $r$-vacua of SQCD.   The dynamical symmetry breaking and confinement are linked to each other (the former is induced by the $Q$ condensates, which in turn, is triggered by the $Q_{0}$ condensation which is the order parameter of confinement), but are not described by one and the same type of condensates.

 \section{Summary}

We have studied several other theories \cite{noi2New} having a similar strongly critical SCFT  in the infrared. 
\begin{description}
  \item[(i)]  Colliding $r$-vacua \footnote{This occurs at the critical mass
  $m=\pm2^{\frac{6N-2N_f}{4N-2N_f}}\frac{2N-N_f}{2N}\Lambda $  for general $SU(N)$, $N_{f}$ theory \cite{DiPietro:2011za}.}  of $SU(3)$ theory with $N_{f}=4$: with the $U(4)$ flavor symmetry  unbroken;  
  \item[(ii)]   Singular $r=2$ vacua \cite{noi2}  of $SU(4)$, $N_{f}=4$ theory, with symmetry breaking $U(4) \to  U(2)\times U(2)$; 
    \item[(iii)]    $SO(2N+1)$, $N_{f}=1$ theory with symmetry breaking $ USp(2)=SU(2)  \to  U(1)$; 
        \item[(iv)]    $SO(2N)$, $N_{f}=2$ theory, with symmetry breaking $ USp(4) \to  U(2)$.       
\end{description}
 In all these cases confinement and flavor symmetry breaking have been found to be appropriately described by the GST dual variables. As in the case of the $USp(4)$ theory discussed in some detail above, the picture of confinement and dynamical symmetry breaking  in these systems seems be somewhat subtler  than is expected in  a straightforward dual superconductivity mechanism.  This might suggest a new direction in improving our understanding of quark confinement in the real-world  QCD.  

  For the moment our analyses are restricted to special cases where the GST description simplifies particularly, i.e., either to local ones or to those
which can be replaced by another simple system, containing a local but asymptotically free sector (to generate effectively the structure \cite{noi2New}  of the 
GST system, Fig.~\ref{GST}).  It is to be seen whether such a procedure can be applied to more general systems with arbitrary gauge group and for general values of $N_{f}$.

 \section*{Acknowledgement}
This talk is based mainly on the two papers with Simone Giacomelli \cite{noi2,noi2New}. The author thanks the organizer of the Workshop  SCGT12 (Nagoya), K. Yamawaki, for inviting him to present these results in such a stimulating meeting.

\bibliographystyle{ws-procs975x65}
\bibliography{ws-pro-sample}

\end{document}